\documentstyle[graphicx]{mn}

\newif\ifAMStwofonts
\AMStwofontstrue

\def\ngc{{NGC 4051}}

\def\xmm{{\it XMM-Newton}}
\def\chandra{{\it Chandra}}

\def\et{{et al.\ }}

\def\rosat{{\it ROSAT}}
\def\ginga{{\it GINGA}}

\def\asca{{\it ASCA}}
\def\sax{{\it BeppoSAX}}
\def\xte{{\it RXTE}}

\newcommand{\ls}{\mathrel{\hbox{\rlap{\hbox{\lower4pt\hbox{$\sim$}}}\hbox{$<$}}}}
\newcommand{\gs}{\mathrel{\hbox{\rlap{\hbox{\lower4pt\hbox{$\sim$}}}\hbox{$>$}}}}

\def\arcs{{\hbox{$^{\prime\prime}$}}}

\def\Msun{\hbox{$\rm ~M_{\odot}$}}

\def\H0{{\rm ~km~s^{-1}~Mpc^{-1}}}

\def\et{{et al.}}

\def\deg{^\circ}

\title[X-ray spectrum of \ngc]
        {Exploring the complex X-ray spectrum of \ngc.}
\author[K.A.Pounds \et]
        {K.A.Pounds$^{1}$,
	J.N.Reeves$^{2}$,
	A.R.King$^{1}$ and 
	K.L.Page,$^{1}$ 
	\\
$^{1}$ Department of Physics and Astronomy, University of Leicester,
Leicester, LE1 7RH, UK\\
$^{2}$ Laboratory for High Energy Astrophysics, NASA Goddard Space Flight Center, Greenbelt, MD 20771, USA\\}

\date{Accepted ; Submitted }
\pagerange{\pageref{firstpage}--\pageref{lastpage}}
\pubyear{2003}
\begin{document}
\maketitle
\label{firstpage}

\begin{abstract} Archival \xmm\ data on the nearby Seyfert galaxy \ngc, taken in relatively high and low flux states,
offer a unique opportunity to explore the complexity of its X-ray spectrum. We find the hard X-ray band to be
significantly affected by reflection from cold matter, which can also explain a non-varying, narrow Fe K fluorescent line.
We interpret major differences between the high and low flux hard X-ray spectra in terms of the varying ionisation (opacity)
of a substantial column of outflowing gas. An emission line spectrum in the low flux state indicates an extended
region of photoionised gas. A high velocity, highly ionised outflow seen in the high state spectrum can replenish the gas in the
extended emission region over $\sim$$10^{3}$ years, while having sufficient kinetic energy to contribute significantly to the 
hard X-ray continuum. 
\end{abstract}

\begin{keywords}
galaxies: active -- galaxies: Seyfert: general -- galaxies:
individual: NGC 4051 -- X-ray: galaxies
\end{keywords}

\section{Introduction}

The additional sensitivity of \xmm\ and \chandra\ has emphasised the complexity in the X-ray spectra of AGN.
While there is broad agreement that the X-ray emission is driven by accretion onto a supermassive black hole, the
detailed emission mechanism(s) remain unclear. Significant complexity - and diagnostic potential - is introduced by
reprocessing of the primary X-rays in surrounding matter. Scattering and fluorescence from dense matter in the putative
accretion disc has been recognised as a major factor in modifying the observed X-ray emission of bright Seyfert galaxies
since its discovery 13 years ago (Nandra \et\ 1989, Pounds \et\ 1990). Additional modification of the observed X-ray
spectra arises by absorption in passage through ionised matter in the line of sight to the continuum X-ray source. The high
resolution X-ray spectra obtained with \xmm\ and \chandra\ have shown the considerable complexity of this `warm absorber'
(eg Sako \et\ 2001, Kaspi \et\ 2002), including recent evidence for high velocity outflows (eg Chartas \et\ 2002, Pounds \et\ 
2003a,b; Reeves \et\ 2003) which constitute a significant component in the mass and energy budgets of
those AGN.  In this paper we report on the spectral analysis of two \xmm\ observations of the bright, nearby Seyfert 1
galaxy \ngc\ taken from the \xmm\ data archive. We find further support for the suggestion made in an early survey of
\xmm\ Seyfert spectra (Pounds and Reeves 2002), that the full effects of ionised absorption in AGN have often been 
underestimated.

\ngc\ is a low redshift ($z=0.0023$) narrow line Seyfert 1 galaxy, which has been studied over much of the history of X-ray
astronomy. Its X-ray emission often varies rapidly and with a large amplitude (Lawrence \et\ 1985,1987), occasionally lapsing into
extended periods of extreme low activity (Lamer \et\ 2003). When bright, the broad band X-ray spectrum of \ngc\ appears typical of
a Seyfert 1 galaxy, with a 2--10 keV continuum being well represented by a power law of photon index $\Gamma$ $\sim$1.8--2, with a
hardening of the spectrum above $\sim$7 keV being attributable to `reflection' from `cold', dense matter, which might also be the
origin of a relatively weak Fe K emission line (Nandra and Pounds 1994). However, \ngc\ also exhibits strong spectral variability,
apparently correlated with source flux. The nature of this spectral variability has remained controversial since the \ginga\ data
were alternatively interpreted as a change in power law slope (Matsuoka \et\ 1990) and by varying partial covering of the
continuum source by optically thick matter (Kunieda \et\ 1992).

Later \rosat\ observations provided good evidence for a flux-linked variable ionised absorber, and for a `soft excess' below $\sim$1
keV (Pounds \et\ 1994, McHardy \et\ 1995,  Komossa and Fink 1997). Extended \asca\ observations led Guainazzi \et (1996) to report a
strong and broad Fe K emission line (implying reflection from the inner accretion disc), and a positive correlation of the hard power
law slope with X-ray flux. A 3-year monitoring campaign of \ngc\ with \xte, including a 150-day extended low interval in 1998, produced
clear evidence for the cold reflection component (hard continuum and narrow 6.4 keV Fe K line) remaining constant, while again finding
the residual power law slope to steepen at higher X-ray fluxes (Lamer \et\ 2003). More surprisingly, a relativistic broad Fe K line
component was found to be always present, even during the period when the Seyfert nucleus was `switched off' (Guainazzi \et\ 1998,
Lamer \et\ 2003). One other important contribution to the extensive X-ray literature on \ngc\ came from an early \chandra\ observation
which resolved two X-ray absorption line systems, with outflowing velocities of $\sim$2300 and $\sim$600 km s$^{-1}$, superimposed on a
continuum soft excess with significant curvature (Collinge \et 2001). Of particular interest in the context of the present analysis,
the higher velocity outflow is seen in lines of the highest ionisation potential. The \chandra\ data also show an unresolved Fe K
emission line at $\sim$6.41 keV (FWHM $\leq$2800 km s$^{-1}$).

In summary, no clear picture emerges from a review of the extensive data on the X-ray spectrum of \ngc, with the spectral variability
being (mainly) due to a strong power law slope - flux correlation, or to variable absorption in (a substantial column of) ionised
matter. Support for the former view has recently come from a careful study of the soft-to-hard flux ratios in extended
\xte\ data (Taylor \et\ 2003), while the potential importance of absorption is underlined by previous spectral fits to \ngc\ requiring
column densities of order $\sim$$10^{23}$ cm$^{-2}$ (eg Pounds \et\ 1994, McHardy \et\ 1995). 

Given these uncertainties we decided to extract 
\xmm\ archival data on \ngc\ in order to explore its spectral complexities. After submission of the present paper, an independent analysis
of the 2002 November EPIC pn data by Uttley \et\ (2003) was published on astro-ph, reaching different conclusions to those we find. We comment briefly 
on these
alternative descriptions of the spectral variability of \ngc\ in Section 9.4.

\section{Observation and data reduction}

\ngc\ was observed by \xmm\ on 2001 May 16/17 (orbit 263) for $\sim$117 ksec, and again on 2002 November 22 (orbit 541) for $\sim$52
ksec. The latter observation was timed to coincide with an extended period of low X-ray emission from \ngc. These data are now public
and have been obtained from the \xmm\ data archive. X-ray data are available in both observations from the EPIC pn (Str\"{u}der \et
2001) and MOS2 (Turner \et\ 2001) cameras, and the Reflection Grating Spectrometer/RGS (den Herder \et\ 2001). The MOS1 camera was also
in spectral mode in the 2002 observation. Both EPIC cameras were used in small window mode in the first observation, together with the
medium filter, successfully ensuring negligible pile-up. The large window mode, with medium filter, was used in the second, low flux
state observation. The X-ray data were first screened with the latest XMM SAS v5.4 software and events
corresponding to patterns 0-4 (single and double pixel events) were selected for the pn data and patterns 0-12 for MOS1 and MOS2. A low
energy cut of 300 eV was applied to all X-ray data and known hot or bad pixels were removed. We extracted EPIC source counts within a
circular region of 45\arcs\ radius defined around the centroid position of \ngc, with the background being taken from a similar region,
offset from but close to the source. The net exposures available for spectral fitting from the 2001 observation were 81.7 ksec (pn),
103.6 ksec (MOS2), 114.3 ksec (RGS1) and 110.9 ksec (RGS2). For the 2002 observation the final spectral data were of 46.6 ksec (pn),
101.9 ksec (MOS1and 2), 51.6 ksec (RGS1) and 51.6 ksec (RGS2). Data were then binned to a minimum of 20 counts per bin, to facilitate
use of the $\chi^2$ minimalisation technique in spectral fitting.  Spectral fitting was based on the Xspec package (Arnaud 1996).  All
spectral fits include absorption due to the \ngc\ line-of-sight Galactic column of $N_{H}=1.32\times10^{20}\rm{cm}^{-2}$ (Elvis \et\
1989). Errors are quoted at the 90\% confidence level ($\Delta \chi^{2}=2.7$ for one interesting parameter).    

We analysed the broad-band X-ray spectrum of \ngc\ integrated over the separate \xmm\ observations, noting the mean flux levels were
markedly different, and perhaps representative of the `high state' and `low state' X-ray spectra of this Seyfert galaxy. [In fact the
2001 May X-ray flux is close to the historical mean for \ngc, but we will continue to refer to it as the `high state' for convenience].
To obtain a first impression of the spectral change we compare in figure 1 the background-subtracted spectra from the EPIC pn camera for
orbits 263 and 541. The same comparison for the EPIC MOS2 data (not shown) is essentially identical. From $\sim$0.3--3 keV the spectral
shape is broadly unchanged, with the 2001 flux level being a factor $\sim$5 higher. From $\sim$3 keV up to the very obvious emission
line at $\sim$6.4 keV the flux ratio decreases, indicating a flatter continuum slope in the low state spectrum over this energy band. On
this simple comparison the $\sim$6.4  keV emission line appears essentially unchanged in energy, width and photon flux. We will defer a
more detailed comparison of the `high' and `low' state data until Section 5, after first modelling the individual EPIC spectra.

\begin{figure}                                                          
\centering                                                              
\includegraphics[width=6.3cm, angle=270]{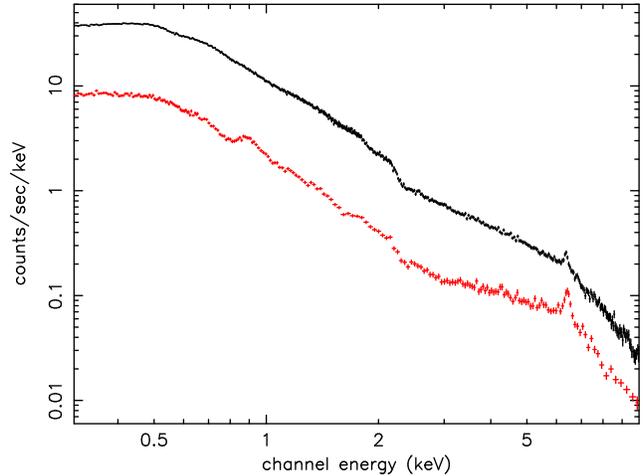}                     
\caption                                                                
{Background-subtracted EPIC pn data for the 2001 May (black) and 2002 November (red) observations of \ngc}      
\end{figure}

\section{High state EPIC spectrum} 

\subsection{Power law continuum}

We began our analysis of the EPIC data for 2001 May in the conventional way by fitting a power law over the hard X-ray
(3--10 keV) band, thereby excluding the more obvious effects of soft X-ray emission and/or low energy absorption. This fit yielded
a  photon index of $\Gamma$$\sim$1.85 (pn) and $\Gamma$$\sim$1.78 (MOS), but the fit was poor with significant residuals.
In particular the presence of a narrow emission line near 6.4 keV, and increasing positive residuals above 9 keV (figure 2),
suggested the addition of a cold reflection component to refine the continuum fit, which we then modelled with PEXRAV in
Xspec (Magdziarz and Zdziarski 1995). Since the reflection component was not well constrained by the continuum fit, we
left free only the reflection factor R (= $\Omega$/2$\pi$, where $\Omega$ is the solid angle subtended at the source),
fixing the power law cut-off at 200 keV and disc inclination at 20$\deg$, with all abundances solar. The outcome was an
improved fit, with $\Delta$$\chi^{2}$ of 40 for R = 0.8$\pm$0.2. The power law index $\Gamma$ increased by 0.1 for both
pn  and MOS fits. In all subsequent fits we then set R = 0.8 (compatible with the strength of the 6.4 keV emission
line).  Based on this broad band fit we obtained a 2-10 keV flux for the 2001 May observation of \ngc\ of
$2.4\times10^{-11}$~erg s$^{-1}$ cm$^{-2}$  corresponding to a 2-10 keV luminosity of $2.7\times 10^{41}$~erg s$^{-1}$ ($
H_0 = 75 $~km\,s$^{-1}$\,Mpc$^{-1}$).

\begin{figure}                                                          
\centering                                                               
\includegraphics[width=6.3 cm,angle=270]{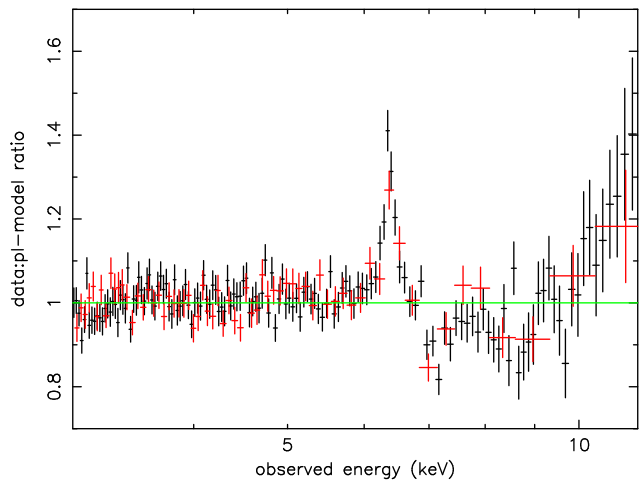}                      
\caption {Ratio of data to power law fits over the 3--10 keV band for the pn (black) and MOS 
(red) spectra in the high state 2001 May
 observation of \ngc.}
\end{figure}

\subsection{Fe K emission and absorption}

The power law plus reflection continuum fit at 3--10 keV leaves several residual features in both pn and MOS data, the significance
of which are indicated by the combined $\chi^{2}$ of 2068 for 1740 degrees of freedom (dof).  Visual examination of figure
2 shows, in particular, a narrow emission line near 6.4 keV and evidence of absorption near $\sim$7 keV and between
$\sim$8--9 keV.

To quantify these features we then added further spectral components to the model, beginning with a  gaussian emission
line with energy, width and equivalent width as free parameters. This addition improved the 3--10 keV fit, to
$\chi^{2}$/dof of 1860/1735, with a line energy (in the AGN rest frame) of 6.38$\pm$0.01 keV (pn) and 6.42$\pm$0.03 keV
(MOS), rms width $\leq$60 eV and line flux of 1.6$\pm$0.4 $\times10^{-5}$~photon s$^{-1}$ cm$^{-2}$ (pn) and 1.4$\pm$0.6
$\times10^{-5}$~photon  s$^{-1}$ cm$^{-2}$ (MOS), corresponding to an equivalent width (EW) of 60$\pm$15 eV. 

Next, we fitted the most obvious absorption feature near 7 keV with a gaussian shaped absorption line, again with energy,
width and equivalent width free. The best-fit observed line energy was 7.15$\pm$0.05 keV (pn) and 7.05$\pm$0.05 keV (MOS) in the
AGN rest-frame, with an rms width of 150$\pm$50 eV, and an EW of 100$\pm$20 eV. The addition of this gaussian absorption
line gave a further highly significant improvement to the overall fit, with $\chi^{2}$/dof = 1802/1730. Fitting the less
compelling absorption feature at $\sim$8--9 keV with a second absorption line was not statistically significant. However,
an absorption edge did improve the fit to $\chi^{2}$/dof = 1767/1728, for an edge energy of 8.0$\pm$0.1 keV and optical
depth 0.15$\pm$0.05. 

In summary, the 3-10 keV EPIC data from the high state 2001 May observation of \ngc\ is dominated by a power law continuum, with a
photon index (after inclusion of cold reflection plus an emission and absorption line) of 1.90$\pm$0.02 (pn) and
1.84$\pm$0.02 (MOS).  The narrow emission line at $\sim$6.4 keV is compatible with fluorescence from the same cold
reflecting matter, while - if identified with resonance absorption of FeXXVI or FeXXV - the $\sim$7.1 keV line implies a
substantial outflow of highly ionised gas. We find no requirement for the previously reported strong, broad Fe K emission
line, the formal upper limit for a line of initial energy 6.4 keV being 70 eV.  

\subsection{Soft Excess}

Extending the above 3--10 keV continuum spectral fit down to 0.3 keV, for both pn and MOS data, shows very clearly
(figure 3) the strong soft excess indicated in earlier observations of \ngc. 

To quantify the soft excess we again fitted the combined pn and MOS data, obtaining a reasonable overall fit
with the addition of blackbody continua of kT $\sim$120 and 270 eV, together with absorption edges at
$\sim$0.725 keV ($\tau$$\sim$0.24) and $\sim$0.88 keV ($\tau$$\sim$0.09). Based on this broad band fit we deduced soft
X-ray flux levels for the 2001 May observation of \ngc\ of $2.9\times10^{-11}$~erg s$^{-1}$ cm$^{-2}$  (0.3--1 keV), with
$\sim$61 percent in the blackbody components, and $1.1\times10^{-11}$~erg s$^{-1}$ cm$^{-2}$  (1--2 keV). Combining these
results with the higher energy fit yields an overall 0.3--10 keV luminosity of \ngc\ in the `high' state of $7\times
10^{41}$~erg s$^{-1}$ ($ H_0 = 75 $~km\,s$^{-1}$\,Mpc$^{-1}$).

\begin{figure}                                                          

\centering                                                              
\includegraphics[width=6.3 cm, angle=270]{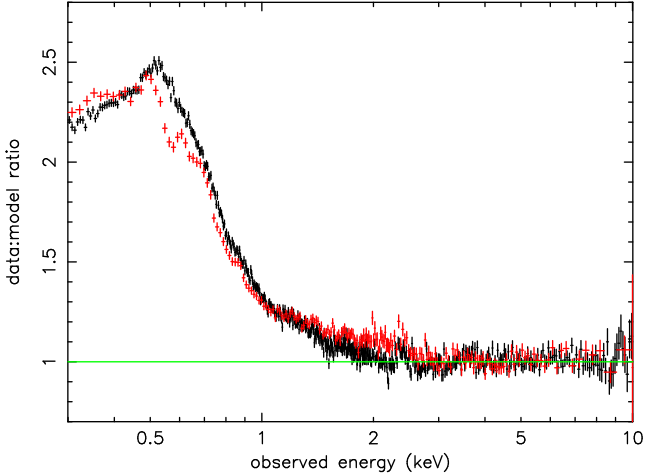}                     
\caption                                                                
{Extrapolation to 0.3 keV of the 3--10 keV spectral fit (detailed in section 3.2) showing the strong soft excess 
in both pn (black) and 
MOS (red) spectra during the 2001 May observation of \ngc.}
\end{figure}

\section{Low state EPIC spectrum}

The above procedure was then repeated in an assessment of the 2002 November EPIC data, when the X-ray flux from \ngc\ was
a factor $\sim$4.5 lower (figure 1).

Fitting the hard X-ray continuum was now more uncertain since the spectrum was more highly curved
in the low flux state (compare figs 4 and 2), making an underlying power law component difficult to identify. To constrain the
fitting parameters we therefore made two important initial assumptions. The first, supported by the minimal change apparent in the
narrow Fe K line, was to carry forward the cold reflection (normalisation and R) parameters from the `high state' spectral fit (in
fact, as noted above, appropriately at a flux level close to the historical average for \ngc ). The second assumption was that the
power law continuum changed only in normalisation, but not in slope (as found in the extended \xmm\ observation of
MCG-6-30-15, Fabian and Vaughan 2003). This is in contrast to the conclusions of Lamer \et\ (2003) but - as we see later - 
is consistent with the
difference spectrum (figure 8), which fits quite well at 3--10 keV to a power law slope of $\Gamma$$\sim$2, while also showing no
significant residual reflection features. 

With these initial assumptions, the 3--10 keV fit to the low state spectrum yielded the data:model ratio shown in figure 4.  A
visual comparison with figure 2 shows a very similar narrow emission line at $\sim$6.4 keV, but with strong curvature to
the underlying continuum, and significant differences in the absorption features above 7 keV. These strong residuals
resulted in a very poor fit at 3--10 keV, with $\chi^{2}$ of 1610/990. We note the spectral curvature in the 3--6 keV band
is reminiscent of an extreme relativistic Fe K emission line; however, since our high state spectrum showed no evidence
for such a feature, and it might in any case be unexpected when the hard X-ray illumination of the innermost accretion
disc is presumably weak, we considered instead a model in which a fraction of the power law continuum is obscured by an
ionised absorber. We initially modelled this possibility with ABSORI in Xspec, finding both the 3--6 keV spectral
curvature and the absorption edge at $\sim$7.6 keV were well fitted with $\sim$60 percent of the power law covered by
ionised matter of ionisation parameter $\xi$(= $L/nr^2$)$\sim$25 and  column density $N_{H}$
$\sim$1.2$\times10^{23}$~cm$^{-2}$. 

The main residual feature was then the narrow Fe K emission line.

\begin{figure}                                                          
\centering                                                                
\includegraphics[width=6.3
cm,angle=270]{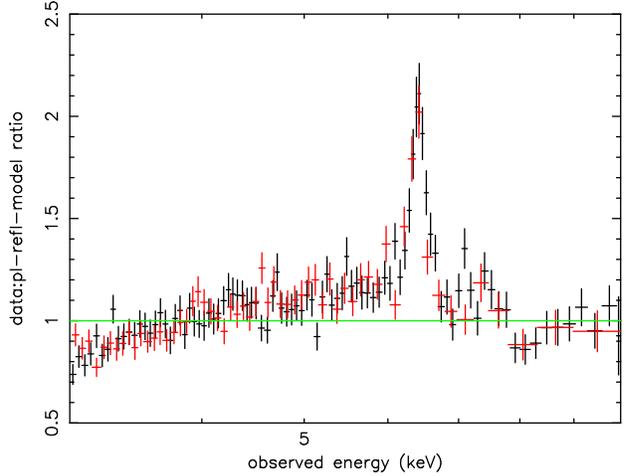}                       
\caption {Ratio of data to power law plus continuum reflection model fit over the 3--10 keV band for
the pn (black) and MOS (red) spectra in the low state 2002 November observation of \ngc.} 
\end{figure}      

\begin{figure}                                                          
\centering                                                                
\includegraphics[width=6.3
cm,angle=270]{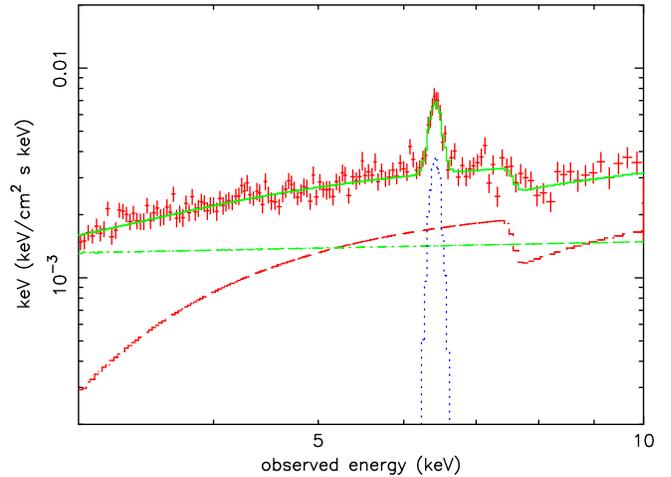}                       
\caption {Partial covering model spectrum fitted over the 3--10 keV band for the 2002 November observation of \ngc. Also shown are
the separate components in the fit: the unabsorbed power law (green), absorbed power law (red) and Gaussian emission line (blue).
See Section 4.1 for details. For clarity only
the pn data are shown.} 
\end{figure}

\subsection{The narrow Fe K emission line}

A gaussian line fit to the emission line at $\sim$6.4 keV in the low state EPIC data was again unresolved, with a mean energy (in the
AGN rest frame) of 6.41$\pm$0.01 keV (pn) and 6.39$\pm$0.02 keV (MOS), and line fluxes of 1.9$\pm$0.3 $\times10^{-5}$~photon s$^{-1}$
cm$^{-2}$ (pn) and 2.0$\pm$0.4 $\times10^{-5}$~photon s$^{-1}$ cm$^{-2}$ (MOS), corresponding to an EW against the unabsorbed power law
component of 500$\pm$75 eV. The important point is that, within the measurement errors, the measured fluxes of the $\sim$6.4 keV
line are the same for the two observations. This lends support to our initial assumption that both EPIC spectra include a `constant'
reflection component, illuminated by the long-term average hard X-ray emission from \ngc. With the addition of this narrow emission line
the overall 3--10 keV fit obtained with the partial covering model was then good ($\chi^{2}$/dof = 1037/1037). Figure 5 illustrates the
unfolded spectrum and spectral components of this fit.

\subsection{Soft Excess}

Extrapolation of the above partial covering 3--10 keV spectral fit down to 0.3 keV shows a substantial soft X-ray excess
remains (figure 6), with a similar relative strength to the power law component seen in the high state data. We note that the `soft excess', 
ie relative to the 
power law component, would have been extremely
strong (data: model ratio$\sim$8) had  we taken the simple power law fit ($\Gamma$$\sim$1.4) to the low state 3--10 keV
data. Extending the partial covering model  to 0.3 keV, with the addition - as in the high state - of a blackbody component of 
kT $\sim$125 eV (the hotter
component was not required), 
gave an initially poor fit ($\chi^{2}$ of 2348 for 1265 dof for the pn data), with a broad deficit in observed flux at
$\sim$0.7-0.8 keV being a major contributor (figure 6). The addition of a gaussian absorption line to the partial
covering model gave a large improvement to the broad-band fit  (to $\chi^{2}$ of 1498 for 1262 dof), for a
line centred at 0.756$\pm$0.003 keV, with rms width 50$\pm$15 eV and EW$\sim$40 eV. We show this complex spectral
fit in figure 7, and comment that the model dependency of unfolded spectra is relatively unimportant in 
illustrating such strong, broad band spectral features. Significantly, the broad-band spectral fit remains substantially inferior
to the similar fit to the high state data. Examination of the spectral residuals shows this is due to additional 
fine structure in the soft band of the low state 
spectrum, structure that is also evident in figures 6 and 7. We examine the RGS data in Section 6 to explore the nature (absorption or
emission) of this structure.

The deduced soft X-ray flux levels for the 2002 November observation of \ngc\ were $6.3\times10^{-12}$~erg s$^{-1}$
cm$^{-2}$  (0.3--1 keV), with $\sim$53 percent in the blackbody component, and $1.8\times10^{-12}$~erg s$^{-1}$ cm$^{-2}$ 
(1--2 keV). Combining these results with a 2-10 keV flux of $5.8\times10^{-12}$~erg s$^{-1}$
cm$^{-2}$ yielded an overall 0.3--10 keV luminosity of \ngc\ in the
`low' state of $1.5\times 10^{41}$~erg s$^{-1}$ ($ H_0 = 75 $~km\,s$^{-1}$\,Mpc$^{-1}$).

\begin{figure}                                                          
\centering                                                               
\includegraphics[width=6.3 cm,angle=270]{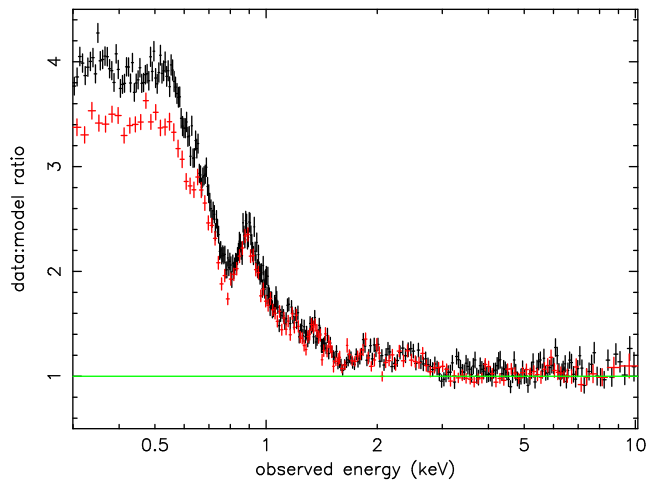}                      
\caption {Partial covering model fits over the 3--10 keV band extended to 0.3 keV, for the pn (black) and MOS (red) 
data from the low state 2002 November observation of \ngc.}
\end{figure}

\begin{figure}                                                          
\centering                                                              
\includegraphics[width=6.3 cm, angle=270]{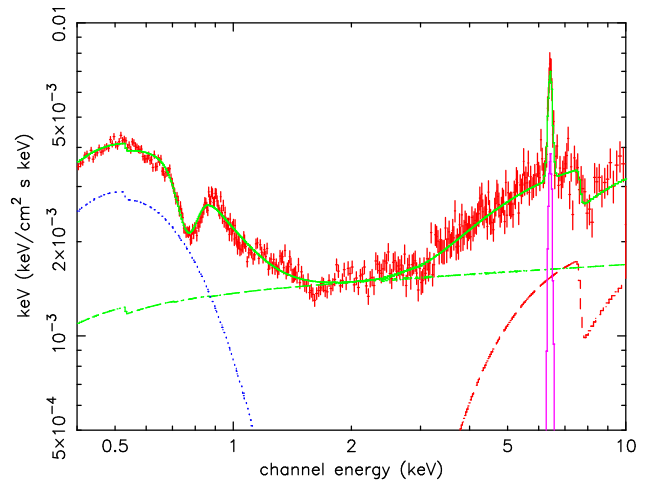}                     
\caption                                                                 
{Extrapolation to 0.3 keV of the 3--10 keV
partial covering fit of fig 5 showing the strong soft excess modelled by a blackbody component (blue), and a broad absorption trough at
$\sim$0.76 keV. For clarity only the pn data are shown.}
\end{figure}      

\section{Comparison of the High and Low state EPIC data}
The above spectral fitting included two important assumptions, that the cold reflection was unchanged between the high and low flux
states, and the variable power law component was of constant spectral index. We now compare the EPIC data for the two observations to
further explore the nature of the spectral change. Figure 8 illustrates the difference spectrum obtained by subtracting the
background-subtracted low state data from the equivalent high state data (corrected for exposure). To improve the statistical
significance of the higher energy points the data were re-grouped for a minimum of 200 counts. The resulting difference
spectrum is compared in figure 8 with a power law fitted at 3--10 keV. Several points are of interest. First, the power law index of the
difference spectrum, $\Gamma$$\sim$2.04 (pn) and $\Gamma$$\sim$1.97 (M2), is consistent with the assumed `constant' value in the
individual spectral fits. Second, the narrow Fe K emission line and high energy data upturn are not seen, supporting our initial
assumption of a `constant' cold reflection component.  The narrow feature observed at $\sim$7 keV corresponds to the absorption line
seen (only) in the high state spectrum, while we shall see in Section 6 that the deficit near 0.55 keV in the MOS data (which has
substantially better
energy resolution there than the pn) is probably explained by a strong and `constant flux' emission line of OVII. Finally, the small peak
near 8 keV can be attributed to the absorption edge shifting to lower energy as the photoionised gas recombines in the reduced continuum
irradiation. 

While the arithmetic difference of two spectra provides a sensitive check for the variability of additive spectral components, a test of the
variability of multiplicative components is provided by the ratio of the respective data sets. Figure 9 reproduces the ratio of the high
and low state data (pn only) after re-grouping to a minimum of 500 counts per bin. From $\sim$0.3--3 keV the flux ratio averages $\sim$5,
as seen in figure 1, falling to higher energies as the mean slope of the low state spectrum hardens. The large positive feature at
$\sim$0.7--0.8 keV is of particular interest, indicating a variable multiplicative component, almost certainly corresponding to enhanced
absorption in the low state spectrum. In fact that feature can be clearly seen in the low state EPIC data in figures 6 and 7. We suggest
the broad excess at $\sim$1--2 keV can be similarly explained by greater absorption affecting the low state spectrum, lending support to
our overall interpretation of the spectral change. Finally, we note that the narrow dip in the ratio plot at $\sim$6.4 keV is consistent
with the Fe K emission line having unchanged flux, but correspondingly higher EW in the low state spectrum.

\begin{figure}                                                          
\centering                                                              
\includegraphics[width=6.3 cm, angle=270]{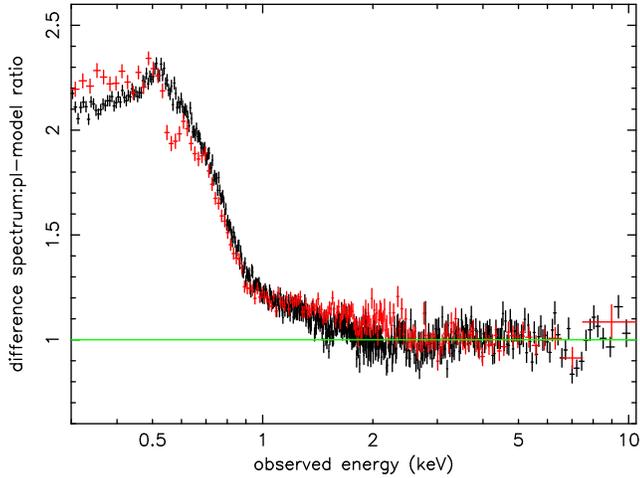}                     
\caption                                                                 
{High minus low state difference spectral data (pn-black, M2-red) compared with a simple power law, as described in Section 5.}
\end{figure}      

\begin{figure}                                                          
\centering                                                              
\includegraphics[width=6.3 cm, angle=270]{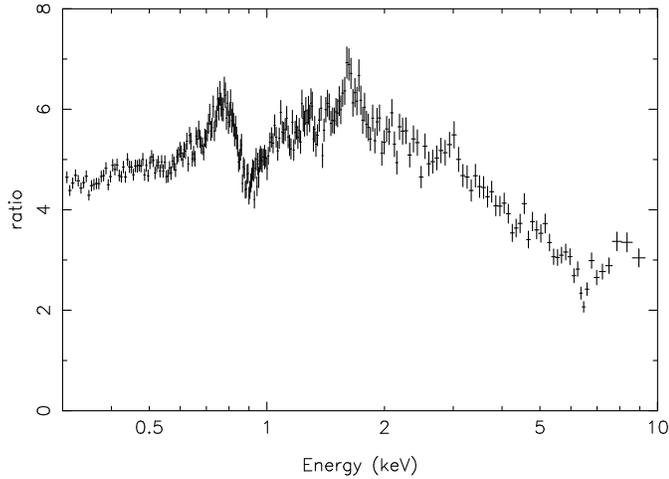}                     
\caption                                                                 
{Ratio of high state to low state spectral data (pn only), as described in Section 5.}
\end{figure}

\section{Spectral lines in the RGS data}

Both EPIC spectra show a strong soft excess, with the low state (2002) spectrum  also having more evidence of fine structure.  To study the
soft X-ray spectra in more detail we then examined the simultaneous \xmm\ grating data for both observations of \ngc. Figures 10 and 11
reproduce the fluxed spectra, binned at 35m\AA, to show both broad and narrow features. The continuum flux level is higher in the 2001 data
(consistent with the levels seen in the EPIC data), with a more pronounced curvature longwards of $\sim$15\AA. Numerous sharp data drops
hint at the presence of many narrow absorption lines. In contrast, the 2002 November RGS spectrum exhibits a lower and flatter continuum
flux, and a predominance of narrow {\it emission} lines.

\begin{figure} 
\centering 
\includegraphics[width=6.3 cm, angle=270]{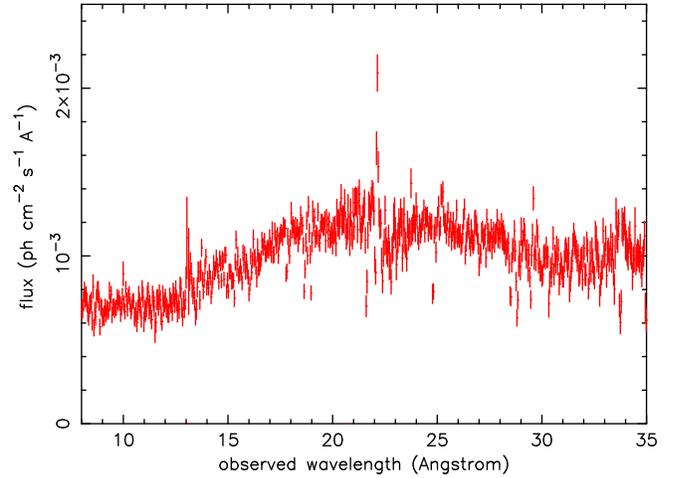} 
\caption {Fluxed RGS spectrum from the \xmm\ observation of \ngc\ in 2001 May.}  
\end{figure}  

\begin{figure} 
\centering 
\includegraphics[width=6.3 cm, angle=270]{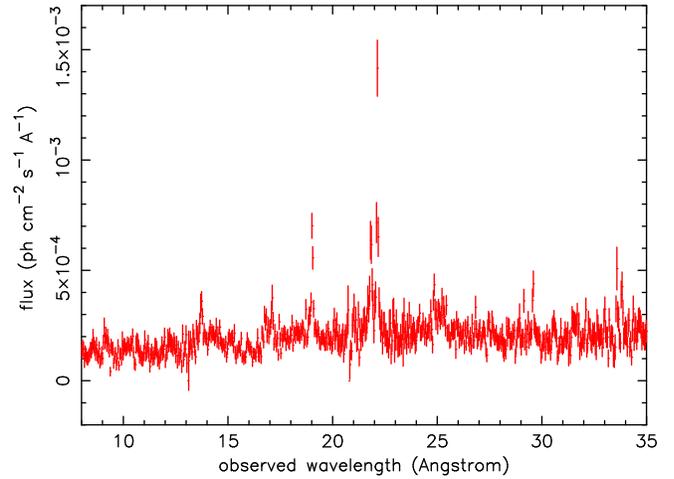} 
\caption {Fluxed RGS spectrum from the \xmm\ observation of \ngc\ in 2002 November.}  
\end{figure}

We began an analysis of each observation by simultaneously fitting the RGS-1 and RGS-2 data with a power law and black body
continuum (from the corresponding EPIC 0.3--10 keV fits) and examining the data:model residuals by eye. For the 2001 May observation the
strongest features were indeed narrow absorption lines, most being readily identified with resonance absorption in He- and H-like ions
of C, N, O and Ne. In contrast, the combined RGS data for the low state data from 2002 November showed a mainly emission line spectrum,
more characteristic of a Seyfert 2 galaxy (eg Kinkhabwala \et 2002). Significantly, the NVI, OVII and NeIX forbidden lines are seen in
both high and low state RGS spectra at similar flux levels.  Taking note of that fact we then analysed the
low state (2002) data first, and subsequently modelled the RGS high-minus-low difference spectrum, to get a truer measure of the absorption
line strengths in the high state (2001) spectrum.

\subsection{An emission line spectrum in the low state data}

To quantify the emission lines in the 2002 spectrum we added gaussian lines to the power law plus blackbody continuum fit in Xspec,
with wavelength and flux as free parameters. In each case the line width was unresolved, indicating a FWHM$\leq$300 km s$^{-1}$.
Details of the 8 strongest lines thereby identified are listed in Table 1.
The statistical quality of the fit was greatly improved by the addition of the listed lines, with a reduction in
$\chi^{2}$ of 251 for 16 fewer dof. When adjusted for the known redshift of \ngc\ all the identified lines are consistent with the
laboratory wavelengths indicating that the emitting gas has a mean outflow (or inflow) velocity of $\leq$200 km s$^{-1}$.

Figure 12 illustrates the OVII triplet, showing the dominant forbidden line and strong intercombination line emission, but no
residual resonance line emission (at 21.6\AA). The line ratios, consistent with those found in the earlier \chandra\ observation
(Collinge \et\ 2001), give a clear signature of a photoionised plasma, with an
electron density $\leq$$10^{10}$cm$^{-3}$ (Porquet and Dubau 2000). A similarly dominant forbidden line in the NVI triplet yields a
density limit  a factor $\sim$10 lower. We note the absence of the OVII resonance emission line may be due to infilling 
by a residual
absorption line of similar strength.

After removal of the emission lines listed in Table 1, several additional emission features (see figure 11) remained. Although narrow  
and barely resolved, the wavelength of these features allows them to be unambiguously identified
with the radiative recombination continua (RRC) from the same He- and H-like ions of C, N, O and (probably) Ne. Table 2 lists the
properties of these RRC as determined by fitting in Xspec with the REDGE model. While the RRC of CV, CVI, NVI and OVII are well
determined, we fixed the other threshold energies at their laboratory values to quantify the measured equivalent widths. What is clear
is that the RRC are very narrow, a combined fit yielding a mean temperature for the emitting gas of kT $\sim$3 eV (T $\sim$$4\times10^{4}$ K). We
note this low temperature lies in a region of thermal stability for such a photoionised gas (Krolik \et\ 1981). Furthermore, the low
temperature indicates collisional ionisation and excitation will be negligible, and radiative recombination should be the dominant
emission process.

Additional constraints on the emitting gas in \ngc\ can be derived by noting that the 2002 November \xmm\ observation took
place some 20 days after the source entered an extended low flux state. Furthermore, the emission line strength of the OVII
forbidden line is essentially the same as when \ngc\ was much brighter in 2001 May. This implies that the emission
spectrum  arises from ionised matter which is widely dispersed and/or of such low density that the recombination time is
$\ga$$2\times10^{6}$s. At a gas temperature of $\sim$$4\times10^{4}$ K, the
recombination time for OVII  is of order 150$(n9)^{-1}$s, where n9 is the number density of the ionised matter in units of
$10^{9}$cm$^{-3}$ (Shull and Van Steenberg 1982). The persistent low state emission would therefore indicate a plasma density 
$\leq$$10^{5}$cm$^{-3}$.

\begin{table*}
\centering
\caption{Principal emission lines identified in the 2002 November RGS spectrum of \ngc. Wavelengths are in Angstroms
and adjusted to the source rest frame and line fluxes are in units of $10^{-5}$~photons cm$^{-2}$ s$^{-1}$.}

\begin{tabular}{@{}lccccc@{}}
\hline
Line &  $\lambda$$_{source}$ & $\lambda$$_{lab}$ &  flux  & EW (eV) \\

\hline
CVI Ly$\alpha$ & 33.75$\pm$0.03 & 33.74 & 4$\pm$1 & 2.5$\pm$0.6 \\
NVI 1s-2p (f) & 29.55$\pm$0.04 & 29.53 & 3.5$\pm$0.5 & 3.2$\pm$0.5  \\
NVII Ly$\alpha$ & 24.79$\pm$0.03 & 24.78 & 3$\pm$0.8 & 3.5$\pm$1 \\
OVII 1s-2p (f) & 22.14$\pm$0.02 & 22.10 & 11$\pm$1.5 & 14$\pm$2  \\
OVII 1s-2p (i) & 21.79 $\pm$0.03 & 21.80 & 5$\pm$1 & 6$\pm$2  \\
OVIII Ly$\alpha$ & 18.99 $\pm$0.03 & 18.97 & 8$\pm$1.3 & 10$\pm$1.5 \\
FeXVII 2p-3s & 17.06 $\pm$0.03 & 17.2 & 2$\pm$0.6 & 5$\pm$2  \\
NeIX 1s-2p (f) & 13.75 $\pm$0.05 & 13.70 & 3$\pm$0.6 & 7$\pm$2 \\
\hline
\end{tabular}
\end{table*}

Assuming a solar abundance of oxygen, with 30 percent in OVII, 50 percent of recombinations from OVIII direct to the
ground  state, and a recombination rate at kT $\sim$3 eV of $10^{-11}$~cm$^{3}$~s$^{-1}$ (Verner and Ferland 1996), we deduce an emission measure for
the forbidden  line flux of order $2\times$$10^{63}$cm$^{-3}$. That corresponds to a radial extent of $\ga$$3\times10^{17}$cm
for a uniform spherical distribution of photoionised gas at the above density of  $\leq$$10^{5}$cm$^{-3}$. Coincidentally,
the alternative explanation for a constant emission line flux, via an extended light travel time, also requires an
emitting region scale size of $\ga$$10^{17}$cm. We note, furthermore, that these values of particle density and radial distance from
the ionising continuum source are consistent with the ionisation  parameter  derived from our XSTAR fit to the RGS
absorption spectrum (Section 7). The scale of the soft X-ray emitting gas
is apparently much greater than the BLR, for which Shemmer \et\ 2003 find a value of 3.0$\pm$1.5 light days ($\sim$$3-10\times 10^{15}
$cm). In fact it has overall properties, of  density, temperature and velocity
consistent with the NLR in \ngc.

The above emission lines and RRC provide an acceptable fit to the RGS data for the 2002 November observation of \ngc. However a coarse
binning of the data:model residuals (figure 13) shows a broad deficit of flux remaining at $\sim 15-17$\AA. It seems likely that
this feature is the same as that seen in the broad band fits to the EPIC data for 2002 November (Section 4) and tentatively identified
with an unresolved transition array (UTA) from Fe M-shell ions (Behar \et\ 2001). When fitted with a gaussian  absorption line we find
an rms width of $\sigma$= $\sim$30 eV and EW of 25 eV against the low state continuum, consistent with the absorption trough  required
in the partial covering fit to the low state EPIC data (section 4.2).

\begin{figure} 
\centering 
\includegraphics[width=6.3 cm, angle=270]{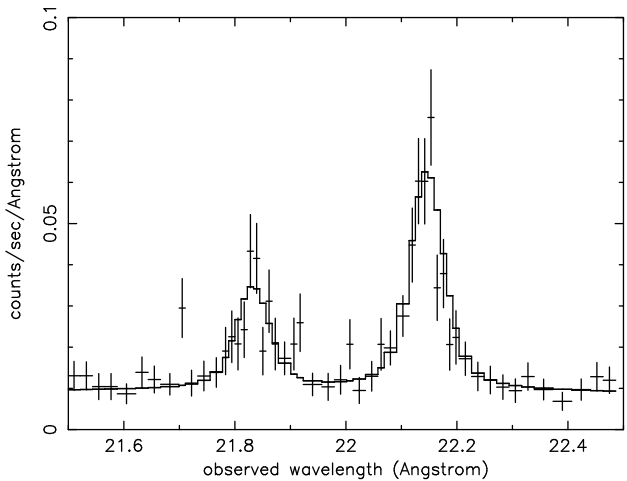} 
\caption {Emission lines dominate the 2002 November RGS data. The OVII triplet is illustrated 
with only the forbidden and intercombination lines clearly visible. The gaussian line fits include only the RGS resolution
showing the emission lines are intrinsically narrow. See Section 6.1 for details.}  
\end{figure}

\begin{figure} 
\centering 
\includegraphics[width=6.3 cm, angle=270]{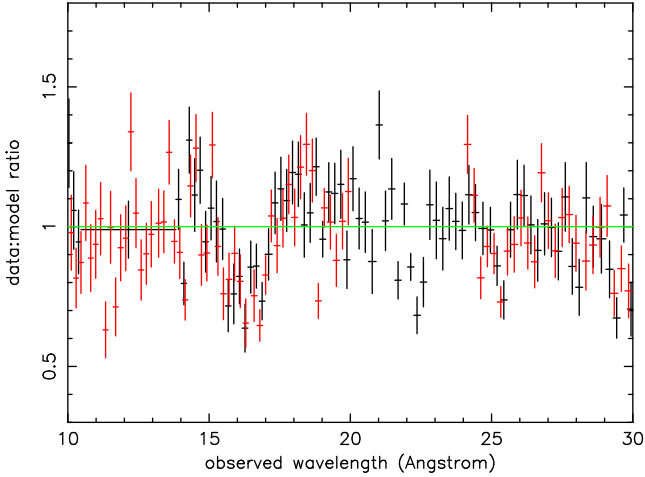} 
\caption {Ratio of the 2002 November RGS data to the emission line and RRC model described in Section 6.1. A broad deficit of
flux at $\sim 15-17$\AA may be attributed to an unresolved transition array (UTA) in weakly ionised Fe.}  
\end{figure}

\begin{table*}
\centering
\caption{Radiative recombination continua identified in the 2002 November RGS spectrum of \ngc. All wavelengths are in 
Angstroms and line fluxes are in units of $10^{-5}$~photons cm$^{-2}$ s$^{-1}$. The threshold wavelengths fixed (f) at the 
laboratory values are as indicated.}

\begin{tabular}{@{}lccccc@{}}
\hline
RRC &  $\lambda$$_{source}$ & $\lambda$$_{lab}$ &  flux & EW (eV) \\

\hline
NeX & 9.13 (f) & 9.11 & 0.3$\pm$0.2 & 6$\pm$4 \\
NeIX & 10.39 (f) & 10.37 & 0.75$\pm$0.3  & 13$\pm$5  \\
OVIII & 14.19 (f) & 14.15 & 1.7$\pm$0.5  &  9$\pm$3 \\
OVII & 16.85 & 16.81 & 3.2$\pm$0.8 & 11$\pm$3 \\
NVI & 22.50 &  22.45 & 1.7$\pm$0.8 & 2$\pm$1 \\
CVI & 25.41 & 25.35 & 5.7$\pm$1 & 7$\pm$1 \\
CV & 31.71 & 31.64 & 2$\pm$1.5 & 3$\pm$2 \\

\hline
\end{tabular}
\end{table*}

\subsection{Absorption lines in the high state difference spectrum.}

The observed wavelengths of the main emission lines in the 2002 spectrum and their equivalent absorption lines in the 2001
spectrum are the same within the resolution of our gaussian line fitting. (At higher resolution the absorption lines appear
to  have a mean outflow velocity of $\sim$500 km s$^{-1}$, while the emission lines are close to the systemic velocity of
\ngc.) Furthermore, from our analysis in Section 6.1 it seems clear that the emission line spectrum
represents an underlying component that responds to some long-term average flux level of the ionising continuum of \ngc.
We therefore first subtracted the 2002 RGS spectrum from the 2001 spectrum with the aim of obtaining a truer measure of
the absorption line strengths in the high state data. Quantifying the main absorption lines by adding gaussian lines
to the corresponding continuum fit then produced the line list in Table 3. 

\begin{table*}
\centering
\caption{Principal absorption lines identified in the 2001 May RGS `difference' spectrum of \ngc. The Ne IX resonance line 
appears to 
be broad and is probably blended
with Fe XIX. Wavelengths are in Angstroms and adjusted to the source restframe}

\begin{tabular}{@{}lcccc@{}}
\hline
Line &  $\lambda$$_{source}$ & $\lambda$$_{lab}$ & EW (mA) \\

\hline
CVI Ly$\alpha$ & 33.67$\pm$0.02 & 33.74 & 170$\pm$30 \\
CVI Ly$\beta$ & 28.42 $\pm$0.02 & 28.47 & 60$\pm$15 \\
CVI Ly$\gamma$ & 26.95$\pm$0.03 & 26.99 & 20$\pm$10 \\
NVI 1s-2p & 28.73$\pm$0.02 & 28.79 & 80$\pm$15  \\
NVII Ly$\alpha$ & 24.73$\pm$0.03 & 24.79 & 100$\pm$25 \\
OVIII Ly$\alpha$ & 18.93 $\pm$0.01 & 18.97 & 100$\pm$15  \\
OVIII Ly$\beta$ & 15.98 $\pm$0.02 & 16.01 & 25$\pm$10  \\
OVII 1s-2p & 21.57 $\pm$0.03 & 21.60 & 60$\pm$15  \\
OVII 1s-3p & 18.62 $\pm$0.03 & 18.63 & 40$\pm$112  \\
OVII 1s-4p & 17.76 $\pm$0.04 & 17.77 & 15$\pm$8  \\
NeIX 1s-2p & 13.43 $\pm$0.03 & 13.45 & 40$\pm$15  \\

\hline
\end{tabular}
\end{table*}

\section{An ionised absorber model fit to the 2001 May \xmm\ data.}

To better quantify the highly ionised matter responsible for the observed absorption features in the `high flux state' spectrum of
\ngc\ and check for physical consistency of the candidate line identifications, including the high energy absorption
features seen above 7 keV in the EPIC data, we next replaced the gaussian absorption lines in the above fits with a model
comprising a grid of photoionised absorbers based on the XSTAR code (Kallman \et\ 1996). We modelled the RGS 
difference spectrum as the best
measure of low ionisation matter; however, in the case of the EPIC data, where we expect the particle density to be higher 
(and recombination time shorter) than in the extended emission region, we modelled the direct high state data. The XSTAR model 
absorbers cover a wide range of
column density and ionisation parameter, with outflow (or inflow) velocities as a variable parameter. All abundant
elements from C to Fe are included with the relative abundances as a variable input parameter. To limit processing time
we assumed a fixed width for each absorption line of 1000 km s$^{-1}$ FWHM. 

We first attempted a fit to the RGS data over the 8.5--35\AA band, where the power law plus blackbody continuum gave
$\chi^{2}$ of 5388 for 4197 dof. The addition of a two-component ionised absorber significantly improved this fit, to $\chi^{2}$ of
4829/4183, with an ionisation parameter log$\xi$  of 2.7$\pm$0.1 and column density of $N_{H}$$\sim$$6\times10^{21}\rm{cm}^{-2}$,
and an ionisation parameter log$\xi$  of 1.4$\pm$0.1 and column density of $N_{H}$$\sim$$2\times10^{21}\rm{cm}^{-2}$. The relative
abundances of C,N,O,Ne, and Fe, tied for both components, were  determined to be 0.5, 0.6,0.35,0.3 and 1.0. The
apparent redshift from the fit was 5.2$\pm$1.4$\times10^{-4}$, indicating an  outflow velocity of $\sim$600 km s$^{-1}$ at the
redshift of \ngc. We note this velocity is probably a lower limit for those lines where a significant emission component has been
subtracted. While the present analysis is only intended to obtain a
rough characterisation of the low ionisation outflow, several points stand out.

A significantly better fit with the 2-component absorber suggests a range of ionisation parameter exists in the outflow, a
point emphasised more strongly when the Fe-K absorption line is added (see below). As noted earlier the strongest observed
lines are all from H- and He-like ions of the lighter metals. The model is seen to fit the the resonance lines of 
OVII and OVIII quite
well (figure 14), perhaps since these strongest lines drive the fit. However, the higher series lines of OVII are noticeably
stronger than in the model, suggesting the resonance line is saturated in the core.

\begin{figure} 
\centering 
\includegraphics[width=6.3 cm, angle=270]{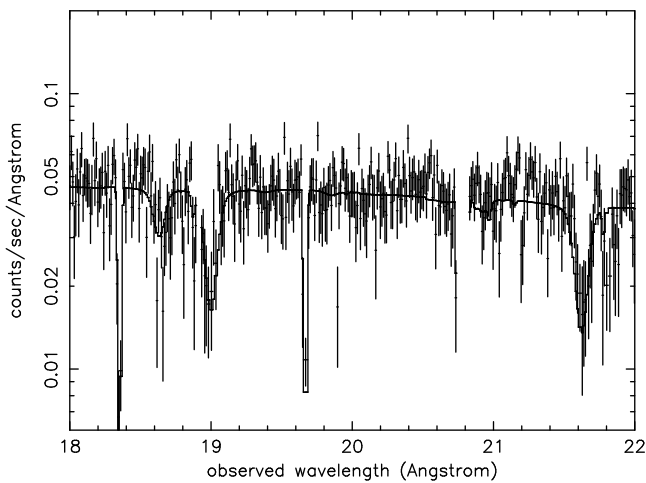} 
\caption {Part of the 2001 May RGS `difference' spectrum fitted with the photoionised absorber model 
described in Section 7.  The main absorption lines shown are, from the right, OVII 1s-2p (21.60\AA), OVIII Ly$\alpha$
(18.97\AA) and OVII 1s-3p (18.63\AA)}  
\end{figure}

The $\sim$7.1 keV absorption line, most conservatively attributed to blue-shifted FeXXVI Ly$\alpha$ resonance absorption, required 
a third, more highly ionised component in the absorbing gas, with ionisation parameter log$\xi$$\sim$ 3.8 and column density of
$N_{H}$$\sim$$2\times10^{23}\rm{cm}^{-2}$. The apparent redshift from the fit was -0.02,  indicating an outflow velocity of
$\sim$6500 km$s^{-1}$ at the redshift of \ngc. The ionisation parameter, and hence column density, are not well constrained by
this single line fit. However, the main uncertainty in the column probably lies on the upside, since the line is apparently broader
than in the XSTAR model fit, while the fitted ionisation parameter is close to that for a peak abundance of FeXXVI. Assuming a cone angle
for the highly ionised outflow of $\pi$ sr, the corresponding outflow mass rate is $\sim$$7\times10^{-3}$\Msun yr$^{-1}$, with an
associated kinetic energy of $\sim$$10^{41}$erg s$^{-1}$, comparable to the high (mean) state hard X-ray luminosity of \ngc. 

Repeating the XSTAR fit for the only likely alternative identification of the $\sim$7.1 keV line, with the resonance (1s-2p)
absorption of Fe XXV, required a lower ionisation parameter for the third component, of log$\xi$=$3.3\pm$0.1 together with a column 
density of
$N_{H}$$\sim$$10^{23}\rm{cm}^{-2}$. The apparent redshift from the fit was then -0.058,  indicating an outflow velocity of
$\sim$16500 km s$^{-1}$ at the redshift of \ngc. Again assuming a cone angle of the highly ionised outflow of $\pi$ sr, the
corresponding outflow mass rate is then $\sim$$5\times10^{-2}$\Msun yr$^{-1}$, with a kinetic energy of $2\times$$10^{42}$erg s$^{-1}$.  

\section{An ionised absorber model fit to the 2002 November EPIC data.}

Implicit in our analysis so far is the constant nature of the extended, low ionisation gas. In contrast, our partial covering fit to
the low state EPIC data requires a substantial column density in a lower ionisation state (than in the high state EPIC
fit), covering $\sim$60 per cent of the hard X-ray continuum source. It seems a reasonable assumption that this 
new absorbing component is formed from the previously highly ionised outflow as the ionisation parameter falls with the reduced
hard X-ray flux. To quantify this change, and obtain an alternative fit to the overall low state 
spectrum of \ngc, we then replaced the ABSORI model of section 5 with XSTAR.

The result supports the ABSORI fitting. Figure 15 illustrates the XSTAR fit to the low state EPIC data, the
parameters being, a power law of $\Gamma$= 1.93$\pm$0.03, $\sim$57 per cent of which is covered by a column of
$N_{H}$$\sim$$3.6\times10^{23}\rm{cm}^{-2}$ and intermediate ionisation parameter log$\xi$=1.4$\pm$0.1. The remaining
power law component, a black body of kT $\sim$125 eV, and a narrow Fe K emission line are covered by an ionised absorber
similar to that fitted to the RGS high state data, with log$\xi$$\sim$ 2.8 and $N_{H}$$\sim$$8\times10^{21}\rm{cm}^{-2}$,
together with a cold column of $\sim$$2\times 10^{20}$ cm$^{-2}$, slightly greater than the Galactic column. It is interesting to note 
that a similar high column
density is required for EPIC fits in both high and low states, with the dramatic spectral change being attributed to part
of that matter recombining in response to the lower hard X-ray flux. Furthermore, when left as a free parameter, the power law slope
in the partial covering fit has a preferred value consistent with that in the high state. In other words, the second important
initial assumption we made in section 5 (in addition to constant cold reflection) is also supported in this analysis.

\begin{figure}                                                          
\centering                                                              
\includegraphics[width=6.3 cm, angle=270]{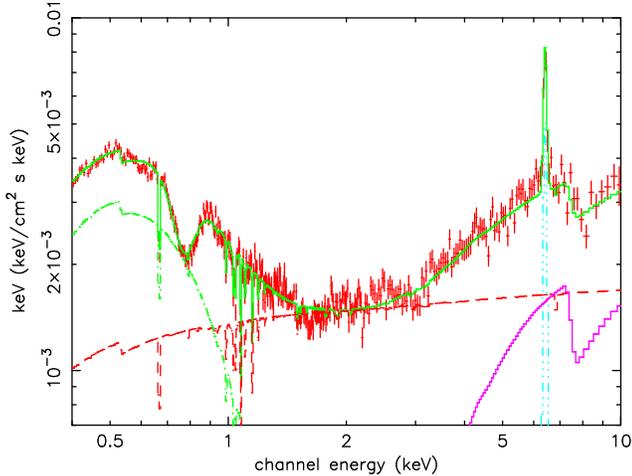}                     
\caption                                                                
{The XSTAR 0.3--10 keV partial covering fit to the low state 2002 November observation of \ngc, showing the strong soft excess 
and a broad absorption trough at $\sim$0.76 keV. Also shown are
the separate components in the fit: the unabsorbed power law (red), absorbed power law (pink), Gaussian emission line (blue) and
blackbody (green).
For clarity only the pn data are shown.}
\end{figure}

\section{Discussion}
\subsection{Hard X-ray emission and re-processing in outflowing gas.}

Given the general acceptance that AGN are powered by accretion onto a supermassive black hole it seems reasonable
that the usually-dominant optical-XUV flux arises as thermal radiation from the accretion disc. However, the origin of the
hard X-ray power law component (and soft X-ray excess) remains less clear, with up-scattering of disc photons in a high temperature `corona' being
a popular mechanism. The strengthening view that viscosity in the disc is largely of magnetic origin offers an appealing
way of transferring accretion energy to the coronal electrons by re-connection in buoyant  magnetic flux. Reprocessing of
hard X-rays in the disc may then explain a major part of the `continuum reflection'  and fluorescent Fe K emission
often seen in AGN spectra.

The most direct evidence given in support of this picture has usually been the rapid, high amplitude X-ray variability (implying a
small emission region) and the broad, skewed profile of the Fe K line, indicating an origin in reflection from the innermost
accretion disc where strong relativistic effects are expected. Recently some doubts have been raised on the wide applicability of
this model. In particular, improved X-ray spectra from \chandra\ and \xmm\ have failed to confirm the relativistic Fe K emission
line in a majority of AGN. Furthermore, new evidence of massive outflows of highly ionised matter in a number of AGN has drawn
attention to the need to take due account of re-processing in overlying (as well as disc) matter. While in the previous cases where
high velocity outflows have been confirmed they appear to be linked to a high (Eddington or super-Eddington) accretion rate, the
evidence for column densities of highly ionised gas in excess of $N_H$$\sim$$10^{23}$cm$^{-2}$ is becoming more common for Seyfert
1 galaxies (eg Bianchi \et\ 2003). Such columns then potentially offer an alternative explanation  to the extreme broad Fe K line,
via partial covering of the power law continuum. In the present analysis of the \xmm\ observations of \ngc\ we have explored the
partial covering alternative, noting that the observation of FeXXVI (or Fe XXV) absorption in the high state requires a similar
column density (of highly ionised gas), which would have recombined as a result of the reduced ionising flux persisting for
$\sim$20 days prior to the 2002 November observation. We conclude that variable opacity in this outflow, responding to the reduced
ionising flux during the extended low state of \ngc, provides a natural explanation of the dramatic change
observed in the broad band X-ray spectrum of \ngc. We suggest that the effects of absorption by line-of-sight ionised gas may have been
generally underestimated in the analysis and interpretation of AGN spectra, and note that a similar explanation was proposed
by Costantini \et\ (2001) in reporting a large scale spectral change in the Seyfert galaxy NGC3516 observed in two \sax\ observations 4
months apart.  

The important detection of a high velocity outflow in the high state spectrum of \ngc\ (given independent support by the recent report 
of an outflow at 4500 km s$^{-1}$ from a second \chandra\ observation; van der Meer \et\ 2003) raises the additional question of what
fraction of the hard X-ray emission may arise, not from the disc/corona, but from shocks in this flow?  We showed in Section 7 that if
the inner flow has a wide cone angle, the associated kinetic energy is comparable with the hard X-ray luminosity in \ngc. If the
trigger for a massive outflow is - as suggested by King and Pounds (2003a) - accretion at or above the Eddington limit, then might this
apply for \ngc ? A recent reverberation measurement (Shemmer \et\ 2003) has indicated the black hole mass in \ngc\ to lie between
$2-10\times 10^{5}$\Msun, an unusually low value (for an AGN), but one supported by a further recent analysis of the X-ray variabilty
(McHardy \et\ 2003).  Such a low mass suggests that in the `typical' bright state of \ngc, as we observed in 2001 May when the total
X-ray  luminosity was $\sim 10^{42} $~erg s$^{-1}$, the bolometric luminosity of \ngc\ might indeed have reached, or exceeded, the
Eddington limit.   

\subsection{Extended photoionised gas in \ngc.} 

The low central continuum flux during the 2002 November observation of \ngc\ allowed the emission spectrum from an extended
photoionised gas to be observed in the RGS data, a rare opportunity to observe this component in a Seyfert 1 galaxy (see also
Turner \et\ 2003 for a similar observation of NGC 3516). The detection of several RRC show the temperature of the gas to be
$\leq$$5\times10^{4}$K, a region of thermal stability, while our analysis of the strong forbidden line emission of OVII provides an
estimate of the (minimum) extent as $3\times$$10^{17}$cm. A corresponding minimum mass for this extended gas envelope is then
$\sim$10\Msun. Assuming a mean outflow velocity of 100 km s$^{-1}$,  the flow time (to reach  $3\times$$10^{17}$cm) is
$\sim$$10^{3}$ years. It is interesting to note that the mass of the extended low ionisation region would be replenished in a
similar timescale if fed by the higher velocity outflow (observed as FeXXVI Ly$\alpha$ absorption at $\sim$7.1 keV).

\subsection{A common outflow} 
Continuing this line of thought we consider how the various components of the overlying gas indicated by the X-ray
spectra of \ngc\ may be parts of the same common outflow. On the (simplest) assumption of a quasi-spherical outflow and 
conservation of mass, we envisage the high velocity/high ionisation 
flow (imprinting absorption features in the Fe K band), degrading into
the intermediate ionisation/lower velocity flow (seen in the RGS absorption lines), before eventually connecting with the larger 
scale/slow moving outflow observed in emission in the low state RGS data.
It is interesting to then speculate that this slowly recombining, low density gas will link into the larger scale outflow resolved in
the [OIII]5007 A line and co-spatial with weak extended radio emission (Christopoulou \et\ 1997).
 
We assume the FeXXVI Ly$\alpha$ interpretation of the 7.1 keV absorption line, at an outflow velocity v of
$\sim$$6\times10^{8}$ cm s$^{-1}$. With an ionising luminosity $\sim$$2\times$$10^{41}$ erg s$^{-1}$ and log$\xi$=3.8, the
product nr$^{2}$v is then $\sim$ $2\times10^{46}$ cm$^{-3}$s$^{-1}$. Conservation of mass in a radial flow maintains
this value to eventually connect with the intermediate and low ionisation components fitted to the RGS high state 
spectra. Assuming each absorbing layer is of thickness 0.3r, the measured column density also allows the mean particle
density (n) and radial distance (r) from the ionising source to be evaluated, yielding n$\sim$$1.2\times10^{10}$ cm$^{-3}$ and
r$\sim$$5\times10^{13}$ cm for the highly ionised flow.

For the intermediate ionisation gas, responsible for most of the observed RGS absorption lines, the measured ionisation
parameter  and column density (together with nr$^{2}$v = $2\times10^{46}$ cm$^{-3}$s$^{-1}$) yield
v$\sim$$5\times10^{7}$cm s$^{-1}$, consistent with the value obtained from the mean line `blueshift', 
together with n$\sim$$10^{6}$ cm$^{-3}$ and
r$\sim$$3\times10^{16}$ cm.

For the lower ionisation component in our XSTAR fit to the RGS absorption lines, the same procedure yields
v$\sim$$2.5\times10^{6}$cm s$^{-1}$, essentially the systemic velocity indicated by the emission line spectrum, together with
n$\sim$$3\times10^{3}$ cm$^{-3}$  and r$\sim$$3\times10^{18}$ cm, again approximately in agreement with the scale deduced
independently from the emission line spectrum.

On this continuous outflow picture the kinetic energy in the initial outflow is largely lost before reaching the intermediate 
ionisation stage, 
and we speculate that this could be due to shocks occurring in the high velocity gas, potentially
contributing a significant part of the hard X-ray luminosity. We note that the derived particle
densities will ensure a rapid response to changing flux in the inner regions of the flow, while the intermediate ionisation
gas 
indicated by the RGS data will already have a recombination time $\ga$ 4 days. 

Important questions that remain from this exploration of the complex X-ray spectrum of \ngc\ include, the mechanism by which the 
high velocity outflow is generated and
the origin of the soft X-ray excess. These questions will be addressed in a separate paper (King and Pounds 2003b).

\subsection{Absorption or spectral pivoting}
Since this paper was initially submitted, an independent analysis of the 2002 November low state data (pn camera only) has been 
accepted for publication in MNRAS (Uttley \et\ 2003). That analysis is guided by the flux-flux plots which, unusually in the case of
\ngc, suggest spectral variability is primarily due to pivoting of the power law spectrum. On that basis Uttley \et\ develop a model for
the low state pn spectrum which contains both variable and constant thermal components to describe the `soft excess'. Their model
also includes a `constant' reflection components and ionised absorption, as does our spectral fit. However, the major difference in the
two approaches is that Uttley \et\ explain the main spectral change by a large change in the power law slope, whereas in the present
paper we emphasise the dominant effects of variable absorption in a substantial column of line-of-sight photoionised gas.

Our analysis has the advantage of including simultaneous high resolution data from the RGS, revealing the presence of an
extended outflow of photoionised gas which we show to be a reasonable continuation of a high density flow at small radii. More direct
evidence for a large column density of ionised gas in line of sight to the X-ray source comes from the detection of a blue-shifted 
FeXXVI Ly$\alpha$ absorption line in high state EPIC data (also not considered in Uttley \et\ 2003). We suggest it is then a natural
outcome for that gas to have recombined during the period of low X-ray emission to give the enhanced absorption implied in our fit to the
low state spectrum of \ngc.  

\section{Summary}

(1) During a typical bright state the X-ray spectrum of the low mass AGN \ngc\ is found to be characteristic of a
Seyfert 1 galaxy, with a canonical power law slope $\Gamma$$\sim$1.9 plus cold reflection (and narrow Fe K emission line),
together with a soft excess superimposed on which is an absorption line spectrum consistent with an outflow of $\sim$500
km s$^{-1}$. An additional spectral feature, not hitherto reported, is a deep absorption line at $\sim$7.1 keV,
indicating a higher velocity, more highly ionised and higher column density outflow component.

(2) A second \xmm\ observation of \ngc, 20 days into a period of unusually faint X-ray emission, exhibits a very different
spectrum. In the soft X-ray band the absorption line spectrum is replaced with an emission line spectrum more
characteristic of a Seyfert 2, which we model as an extended, low density region of photoionised gas. At higher energies
the indicators of cold reflection (including the narrow Fe K emission line) are unchanged, but the continuum exhibits
strong curvature which we interpret in terms of partial covering by a substantial column density of intermediate ionisation.

(3) We conclude that the unusually hard X-ray spectrum of \ngc\ in 2002 November is due to the enhanced visibility of a
`constant' cold reflection component and  
increased
opacity in a substantial column of line-of-sight gas responding to an extended period of low hard X-ray emission.

\section*{ Acknowledgements } The results reported here are based on observations obtained with \xmm, an ESA science mission with
instruments and contributions directly funded by ESA Member States and the USA (NASA). The authors wish to thank Paul O'Brien, Martin
Elvis and Simon Vaughan for valuable input, the SOC and SSC teams for organising the \xmm\ observations  and initial data reduction, and
the referee for a careful and constructive reading of the text. KAP acknowledges the support of a Leverhulme Trust Emeritus Fellowship.

\end{document}